# Constructor Theory of Information

David Deutsch[1] & Chiara Marletto[2]

[1]Centre for Quantum Computation, The Clarendon Laboratory, University of Oxford

[2]Materials Department, University of Oxford

*July 2014*

We present a theory of information expressed solely in terms of which transformations of physical systems are possible and which are impossible – i.e. in constructor-theoretic terms. Although it includes conjectured laws of physics that are directly about information, independently of the details of particular physical instantiations, it does not regard information as an a priori mathematical or logical concept, but as something whose nature and properties are determined by the laws of physics alone. It does not suffer from the circularity at the foundations of existing information theory (namely that information and distinguishability are each defined in terms of the other). It explains the relationship between classical and quantum information, and reveals the single, constructor-theoretic property underlying the most distinctive phenomena associated with the latter, including the lack of in-principle distinguishability of some states, the impossibility of cloning, the existence of pairs of variables that cannot simultaneously have sharp values, the fact that measurement processes can be both deterministic and unpredictable, the irreducible perturbation caused by measurement, and entanglement (locally inaccessible information).

## 1 Introduction

In some respects, information is a qualitatively different sort of entity from all others in terms of which the physical sciences describe the world. It is not, for instance, a function only of tensor fields on spacetime (as general relativity requires all physical quantities to be), nor is it a quantum-mechanical observable.

But in other respects, information does resemble some entities that appear in laws of physics: the theory of computation, and statistical mechanics, seem to refer directly to it without regard to the specific media in which it is instantiated, just as conservation laws do for the electromagnetic four-current or the energy-momentum tensor. We call that the *substrate-independence* of information. Information can also be moved from one type of medium to another while retaining all its properties qua information. We call this its *interoperability* property; it is what makes human capabilities such as language and science possible, as well as biological adaptations that use symbolic codes, such as the genetic code.

Also, information is of the essence in *preparation* and *measurement*, both of which are necessary for testing scientific theories. The output of a measurement is information; the input of a preparation includes information, specifying an attribute with which a physical system is to be prepared.

All these applications of information involve *abstraction*, in that one entity is represented symbolically by another. But information is not abstract in the same sense as, say, the set of all prime numbers, for it only exists when it is physically instantiated. So the laws governing it, like those governing computation – but unlike those governing prime numbers – are laws of physics. In this paper we conjecture what these laws are.

Also, despite being physical, information has a counter-factual character: an object in a particular physical state cannot be said to carry information unless it *could have been* in a different state. As Weaver (1949) put it,

> this word 'information' in communication theory relates not so much to what you *do* say, as to what you *could* say….

The classical theory of information (Shannon 1948) was indeed developed to analyse the physics of communication, where the objective is for a *receiver* to receive a *message* from a *transmitter* through a *medium*. The receiver, transmitter and medium are physical systems, but the message is not. It is information, initially instantiated in the transmitter, then in the medium, then in the receiver. The overall process constitutes a measurement of one of the transmitter's physical variables, representing the message. Essential to Shannon's notion of communication is that a message is one of at least two *possible* messages, which are *distinguishable* by measurement, and that the receiver is able to re-transmit the information to a further receiver while retaining an instance of it. That requires a *non-perturbing* measurement at some stage in the process.

Much of Shannon's theory is about unreliable transmission and measurement, and inefficient representations, and how to compose them into more reliable and efficient ones. But here we are concerned with the fundamental issues that remain even in the limiting case when all error rates have been reduced to their physically possible minima and there is no redundancy in the message being transmitted. In that limit, receiving the message only means *distinguishing* it from all the other possible messages. And in that regard, Shannon's theory is inadequate in two ways.

The first is that it cannot describe information in quantum physics, because certain prohibitions that quantum theory imposes – such as the impossibility of cloning – violate the kind of interoperability that is assumed in Shannon's theory. Consequently the type of information studied by Shannon is now called *classical* information.

The second is that Shannon's theory is about information represented in distinguishable states, but does not specify what distinguishing consists of physically. So, consider the non-perturbing measurement that distinguishes two possible messages $x$ and $y$. It has the following effects in those two cases:

$$\begin{array}{cccccc} \text{message} & \text{receiver} & & \text{message} & \text{receiver} & \\ x & x_0 & \rightarrow & x & x & \\ y & x_0 & \rightarrow & y & y & \end{array} \qquad (1)$$

where $x_0$ is a receptive state of some medium capable of instantiating the outcome $x$ or $y$. But this does not in fact distinguish message $x$ from message $y$ unless the *receiver* states $x$ and $y$ are themselves distinguishable. Therefore (1), considered as a definition of distinguishability, would be circular. Indeed, no existing theory of information provides a non-circular account of what it means for a set of physical states to be mutually distinguishable. The theory that we shall present here does (Section 4).

Likewise *quantum* information theory, as it stands, never gets round to specifying what it is referring to as 'quantum information', nor its relation to classical information. It is not, despite the name, a theory of a new type of information, but only a collection of quantum phenomena that violate the laws of classical information. A new theory of information is needed, within physics but at a deeper level than both quantum theory and Shannon's theory. In this paper we provide that, via *constructor theory* (Deutsch 2013).

Previous attempts to incorporate information at a fundamental level into physics (e.g. Wheeler 1989), or at least into quantum theory (Wootters 1981, Hardy 2001, Clifton *et al.* 2003), have regarded information as being an a priori mathematical or logical concept. Our approach is the opposite, namely that the nature and laws of information follow entirely from those of physics; we are not trying to derive quantum theory from anything. In the theory we present here, the status of information in physics is analogous to that of (say) energy: given the laws of motion of physical objects, neither the concept of energy nor the conservation law for the energy-momentum tensor are necessary for making any prediction from initial data, yet our understanding of the physical world would be radically incomplete without them. The conservation law *explains* some aspects of motion as consequences of a deeper regularity in nature – which is why we expect as-yet-undiscovered laws of motion to conserve the energy-momentum tensor too; but we don't expect to *derive* new laws of motion from it. It is a *principle* – a law of physics that constrains other laws rather than the behaviour of physical objects directly.

After setting out as much of constructor theory as we shall need (Section 2), we shall begin our search for a deeper theory of information by expressing, in exact, constructor-theoretic terms, the concepts of computation, measurement and classical information that are already assumed, informally, to be instantiated in the physical world (Sections 3-5). Then, in Section 6, we express the regularities that are informally associated with *classical* information as exact, purely constructor-theoretic principles of physics – which turn out to be elegant and natural. In particular, we express prediction and testing in constructor-theoretic terms. In Section 7 we introduce *superinformation* media as information media on which certain tasks, with a natural constructor-theoretic definition, are *impossible*. In Section 8 we show that the most distinctive qualitative features of quantum information follow from the impossibility of those tasks. In other words, quantum information is an instance of superinformation.

## 2 Constructor Theory

The laws of constructor theory are all principles, which we conjecture are obeyed by all other laws of physics, so we shall call theories describing those, *subsidiary* theories. Principles, being laws about other laws, do not make direct assertions about the outcomes of measurements. They are nevertheless experimentally testable: a principle *P* is refuted if some law violating *P* survives experimental tests while all rival laws conforming to *P* are refuted[1].

The *basic principle of constructor theory* is that

> **I.** All other laws of physics are expressible entirely in terms of statements about which physical transformations are possible and which are impossible, and why.

This is in contrast with the *prevailing conception* of fundamental physics, which seeks to explain the world in terms of initial conditions and laws of motion, and whose basic dichotomy is therefore between what happens and what does not.

Unlike Shannon's theory, constructor information theory makes no mention of probability. Indeed, principle **I** rules out any reference to probability in fundamental laws of physics. So all 'generalised probabilistic theories' (e.g, Barrett 2007), in which probabilities interpolate between possible and impossible (or between happening and not happening), can at most be descriptions at an emergent level. For a discussion of the emergence of probability in deterministic quantum theory, see Deutsch (1999) and Wallace (2003).

---

[1] For example, early observations of beta decay satisfied the principle of the conservation of energy under the assumption that neutrinos were emitted, but violated it under the assumption that no undetected particle was emitted. The conservation law would have been refuted if the neutrino theory had failed experimental tests while some testable explanation predicting the destruction of energy survived. The supposed principle of parity invariance was refuted in just that way.

Constructor theory describes the world in terms of transformations involving two kinds of systems, playing different roles. One is the object causing the transformation, which we refer to as the *constructor*, and whose defining characteristic is that it remains unchanged in its ability to cause the transformation again. The other is the system being transformed, which may consist of one or more subsystems, the *substrates*:

$$\text{Input attributes of substrates} \xrightarrow{\text{Constructor}} \text{Output attributes of substrates}, \quad (2)$$

where the constructor and the substrates jointly are isolated.

By *'attribute'* we mean anything about the substrate that can possibly be changed by a physical process. We represent it formally as the set of states in which the substrate has that attribute.

An *intrinsic* attribute of a substrate is one that does not refer to any other specific substrate. For example, the distance $x$ if an object $\mathbf{S}_1$ from another object $\mathbf{S}_2$ is an intrinsic attribute of the combined system $\mathbf{S}_1 \oplus \mathbf{S}_2$ but not of either $\mathbf{S}_1$ or $\mathbf{S}_2$. In quantum theory, 'entangled with each other' is a possible intrinsic attribute of a pair of qubits; having a particular density operator is an intrinsic attribute of a system, while the rest of its quantum state (see below) describes entanglement relationships with other systems.

Any set of disjoint attributes, we shall call a physical *variable*. Whenever a substrate is in a state with attribute $x \in X$ where $X$ is a variable, $X$ is *sharp*, with the *value x*.

That individual physical systems (and not just the entire physical world) *have* states, attributes and variables in this sense is guaranteed by Einstein's (1949) *principle of locality* (Einstein 1949), which has a precise expression in constructor-theoretic form:

**II.** There exists a mode of description such that the state of the combined system $\mathbf{S}_1 \oplus \mathbf{S}_2$ of any two substrates $\mathbf{S}_1$ and $\mathbf{S}_2$ is the ordered pair $(x,y)$ of the states $x$ of $\mathbf{S}_1$ and $y$ of $\mathbf{S}_2$, and any construction undergone by $\mathbf{S}_1$ and not $\mathbf{S}_2$ can change only $x$ and not $y$.

In quantum theory, the Heisenberg picture is such a mode of description (see Deutsch & Hayden 2000)[1]. Principle **II** rules out, for example, non-linear modifications of the Schrödinger equation.

The basic entities of constructor theory are specifications of only the input-output pairs in (2), with the constructor abstracted away:

$$\text{Input attributes of substrates} \longrightarrow \text{Output attributes of substrates}.$$

We call these *construction tasks*, or *tasks* for short. In general, a task $\mathfrak{A}$ is a *set* of ordered pairs of intrinsic attributes of some substrates:

$$\mathfrak{A} = \{x_1 \to y_1,\ x_2 \to y_2, ...\}.$$

We call the $\{x_i\} = \text{In}(\mathfrak{A})$ the *legitimate input attributes* of $\mathfrak{A}$ and the $\{y_i\} = \text{Out}(\mathfrak{A})$ its *legitimate output attributes*. The *transpose* of a task $\mathfrak{A} = \{x_1 \to y_1, x_2 \to y_2,...\}$ is $\mathfrak{A}^{\sim} = \{y_1 \to x_1, y_2 \to x_2,...\}$.

A constructor is *capable of performing* a task $\mathfrak{A}$ if, whenever presented with substrates having an attribute in $\text{In}(\mathfrak{A})$, it delivers them with one of the corresponding attributes from $\text{Out}(\mathfrak{A})$ (regardless of what it does if the substrate is in any other state).

Tasks may be composed into networks to form other tasks, as follows. The *parallel composition* $\mathfrak{A} \otimes \mathfrak{B}$ of two tasks $\mathfrak{A}$ and $\mathfrak{B}$ is the task whose net effect on a composite system $\mathbf{M} \oplus \mathbf{N}$ is that of performing $\mathfrak{A}$ on $\mathbf{M}$ and $\mathfrak{B}$ on $\mathbf{N}$. When $\text{Out}(\mathfrak{A}) = \text{In}(\mathfrak{B})$, the *serial composition* $\mathfrak{BA}$ is the task whose net effect is that of performing $\mathfrak{A}$ and then $\mathfrak{B}$ on the same substrate. A *regular network* of tasks is a network without loops whose nodes are tasks and whose lines are their substrates, where the legitimate input states at the end of each line are the legitimate output states at its beginning. Loops are excluded because a substrate on a loop is a constructor.

---

[1] Note that the local states (in our sense) of a system **S** in the Heisenberg picture are the observables of **S**; the global 'state vector' in unchangeable. So the controversy about whether the locality of quantum physics conceals residual 'non-locality' (Wallace & Timpson 2007, Deutsch 2012) is not relevant here because principle **II** only requires *changeable* quantities to be local.

It may be that a task $\{(x,z)\to(y,w)\}$ cannot be decomposed into $\{x\to y\}\otimes\{z\to w\}$ because the individual attributes are not intrinsic and therefore the operands of that parallel composition are not valid tasks. However, if $\{(x,z)\to(y,w)\}$ and $\{x\to y\}$ are valid tasks, then $\{z\to w\}$ must be too.

No perfect constructors exist in nature. Approximations to them, such as catalysts or robots, have non-zero error rates and also deteriorate with repeated use. But we call a task $\mathfrak{A}$ *possible* (which we write as $\mathfrak{A}^{\checkmark}$) if the laws of nature impose no limit, short of perfection, on how accurately $\mathfrak{A}$ could be performed, nor on how well things that are capable of approximately performing it could retain their ability to do so again. Otherwise $\mathfrak{A}$ is *impossible* (which we write as $\mathfrak{A}^{\times}$).

Accordingly we must also understand the principle **I** as requiring subsidiary theories to provide a measure of the accuracy with which any approximate constructor they describe performs tasks (including its own maintenance); and provide a meaning to whether an infinite sequence of tasks $\mathfrak{A}_1, \mathfrak{A}_2, \ldots$ on a system $\mathbf{S}\oplus\mathbf{E}$ (where $\mathbf{E}$ might be the environment of $\mathbf{S}$) converges to a limiting task $\mathfrak{A}$ on $\mathbf{S}$ alone.

A *constructor-theoretic statement* is one that refers only to substrates and which tasks on them are possible or impossible – not to constructors. Constructor theory is the theory that the (other) laws of physics can be expressed *without* referring explicitly to constructors.

A task refers to an isolated system of constructor and substrates. But we are also often interested in what is possible or impossible regardless of the resources required. A task $\mathfrak{A}$ is *possible with side-effects*, which we write as $\mathfrak{A}^{\swarrow}$, if $(\mathfrak{A} \otimes \mathfrak{T})^{\checkmark}$ for some task $\mathfrak{T}$ on some generic, naturally occurring substrate (see Section 6).

## 3 Computation

Our theory of information rests on first understanding *computation* in constructor-theoretic terms. This will allow us to express information in terms of computation; not vice-versa as is usually done. This is the key to avoiding the circularity at the foundations of information theory that we described in Section 1.

A *reversible computation* $\mathfrak{C}_\Pi(S)$ is the task of performing a permutation $\Pi$ over some set $S$ of at least two possible attributes of some substrate:

$$\mathfrak{C}_\Pi(S) = \bigcup_{x \in S} \{x \to \Pi(x)\}.$$

For example, swapping two pure quantum states constitutes a reversible computation, and may be a possible task even if they are not orthogonal. It is then natural to define a *computation variable* as a set $S$ of two or more possible attributes for which $\mathfrak{C}_\Pi^{\swarrow}$ for all permutations $\Pi$ over $S$, and a *computation medium* as a substrate with at least one computation variable. (Since side-effects are allowed in the performance of $\mathfrak{C}_\Pi$, this definition does not require *physical* processes to be reversible.)

Note again that in this paper we are not taking computation to be an a priori concept and seeking necessary and sufficient conditions for a physical process to instantiate it (cf. Horseman et al. 2014). We are conjecturing *laws of physics*: objective regularities in nature. These happen to be conveniently expressed in terms of the tasks we have called 'computations' and the property that we shall call 'information'. We think that these correspond reasonably closely to the intuitive concepts with those names, but our claims in no way depend on that being so.

# 4 Information

As we mentioned in Section 1 the intuitive concept of information is associated with that of copying. We shall express this association exactly and without circularity, in terms of computations as defined in Section 3.

We first consider computations involving two instances of the same substrate **S**. The *cloning task* for a set $S$ of possible attributes of **S** is the task

$$\mathfrak{R}_S(x_0) = \bigcup_{x \in S} \{(x, x_0) \to (x, x)\} \qquad (3)$$

on $\mathbf{S} \oplus \mathbf{S}$, where $x_0$ is some attribute with which it is possible to prepare **S** from generic, naturally occurring resources (Section 6 below). This is a generalisation of the usual notion of cloning, which is (3) with $S$ as the set of *all* attributes of **S**. A set $S$ is *clonable* if $\mathfrak{R}_S(x_0)^{\swarrow}$ for some such $x_0$.

An *information variable* is a clonable computation variable. It is then natural to define an *information attribute* as one that is a member of an information variable, and an *information medium* as a substrate that has at least one information variable.

Also, a substrate **S** *instantiates classical information* if some information variable $S$ of **S** is sharp, and if giving it any of the other attributes in $S$ was possible. And the classical information capacity of **S** is the logarithm of the cardinality of its largest information variable. The principle of locality **II** implies the convenient property that the combined classical information capacity of disjoint substrates is the sum of their capacities.

Thus we have provided the purely constructor-theoretic notion of classical information that we promised. But we have emancipated it from its dependence on classical physics, and cured its circularity.

## 5 Measurement

We can now do the same for distinguishability and measurement. A set *X* of possible attributes of a substrate **S** is *distinguishable* if

$$\left(\bigcup_{x\in X}\{x\to\psi_x\}\right)^{\checkmark}, \tag{4}$$

where the $\{\psi_x\}$ constitute an information variable. Since 'information variable' is defined above without reference to distinguishability, this definition is, as promised, not circular.

If a pair of attributes $\{x,y\}$ is distinguishable we shall write $x\perp y$ (and $x\not\perp y$ if not).

If the original substrate continues to exist and the process (4) stores its result in a second, *output substrate* (which must therefore be an information medium), (4) is the condition for the *input variable X* to be *measurable*:

$$\left(\bigcup_{x\in X}\{(x,x_0)\to(y_x,\text{'}x\text{'})\}\right)^{\checkmark}, \tag{5}$$

where the output substrate is initially prepared with a 'receptive' attribute $x_0$. When *X* is sharp, the output substrate ends up with an information attribute '*x*' of an *output variable*, which represents the abstract outcome «it was *x*». (We give quoted labels such as '*x*' to attributes in the output variable corresponding to those in *X*.) Thus, measurement is like cloning a variable (3) except that the output substrate is an information medium rather than a second instance of the cloned substrate.

A constructor is a *measurer* of *X* if there is some choice of its output variable, labelling, and receptive state, under which it is capable of performing (5). Consequently a measurer of *X* is automatically a measurer of a range of other variables because one can interpret it as such by re-labelling its outputs. (Such re-labellings must be possible tasks because they are classical computations on a finite set – see Section 6.) For example, a measurer of X measures any subset of *X*, or any *coarsening* of *X* (a variable whose members are unions of attributes in *X*).

If $y_x \subseteq x$ in (5), the measurement is *non-perturbing*, which is the type of measurement typically needed in computation and communication. It follows from the definition of information variables that the task of measuring them non-perturbatively is always possible.

**6 Conjectured principles of physics bearing on information**

Crucially, the most important properties of information do not follow from the definitions we have given. In this section we seek the constructor-theoretic principles of physics that determine those properties. Of these, perhaps the most fundamental one cannot even be stated in the prevailing conception of fundamental physics, but it has an elegant expression in constructor theory. It is the *interoperability principle*:

> **III.** The combination of two substrates with information variables $S_1$ and $S_2$ is a substrate with information variable $S_1 \times S_2$, where $\times$ denotes the Cartesian product of sets. Note that this principle requires certain interactions to exist in nature. For instance, it would rule out theories of dark matter in which dark information media exist but interactions between the dark and normal sectors do not allow information to be copied arbitrarily accurately between them.

Exploring the properties of distinguishability in more detail leads us to conjecture further such principles. Suppose that all attributes in a variable *X* are pairwise distinguishable – i.e. $(\forall x \in X, \forall y \in X, x \neq y) x \perp y$. It does not follow logically from the definitions that *X* is a distinguishable variable. But we conjecture that in physical reality, it always is. That is because we expect that whenever there is a regularity among observable phenomena in a substrate (such as a set of its attributes being pairwise distinguishable), that is always because the phenomena are related by a unifying explanation – in this case, that they can all be distinguished by measuring some variable. This, too, has an elegant statement as a purely constructor-theoretic principle:

**IV.** If every pair of attributes in a variable *X* is distinguishable, then so is *X*.

And similarly:

**V.** If every state with attribute *y* is distinguishable from an attribute *x*, then so is *y*.

Though we conjecture that **IV** and **V** hold universally in nature, none of our conclusions in Section 8 depends on their being universal. It would suffice if they held only for a special class of substrates. The same also holds for the remaining conjectured principles (**VI**-**VIII**).

Two of these could be considered simplifying assumptions rather than grand conjectures about the nature of reality. But they might well be true, and are certainly good approximations for present purposes: First, since we are concerned with the nature and properties of information, not its long-term future, we assume that unlimited resources are available for conversion into information storage devices. In constructor-theoretic terms we express this as:

**VI.** Any number of instances of any information medium, with any one of its information-instantiating attributes, is preparable from naturally occurring substrates.

We call such substrates, which must therefore exist in unlimited numbers, 'generic resources'. In describing a task, we use the symbol $g$ to represent a suitable generic resource or resources for the task. Thus for every task $\mathfrak{A}$, we have $\mathfrak{A}^{\checkmark} \Rightarrow \{g \to \mathbf{C}_{\mathfrak{A}}\}^{\checkmark}$, where '$\Rightarrow$' denotes implication and $\mathbf{C}_{\mathfrak{A}}$ is some constructor for $\mathfrak{A}$. The assumption **VI** implies that there are generic (approximations to) constructors in nature too. Thus we also have $\mathfrak{A}^{\checkmark} \Leftrightarrow (\exists h)(\mathfrak{A} \times \{g \to h\})^{\checkmark}$.

And finally we assume that unlimited resources are available for information *processing* too. We express this in constructor-theoretic terms as the conjectured *composition principle* (Deutsch 2013):

**VII.** Every regular network of possible tasks is a possible task,

though here we need only assume that it holds for information-processing tasks.

Provided that information exists at all, principles **III**, **VI** and **VII** imply that for every function *f* from a finite set of integers to itself, the task of computing *f* is possible.

*Measurement of non-sharp variables*

We have defined measurement of a variable *X* by the measurer's effect (5) when *X* is sharp. When we refer to a 'sharp' output we shall mean that the output variable is sharp. But the term 'measurement' is also used to describe cases where *X* is not sharp. In such cases the output variable need not be sharp either. Exploring these will lead us to another constructor-theoretic principle about information.

We first define a convenient tool, the *bar* operation: let *x* be any attribute; $\bar{x}$ ('x-bar') is the union of all attributes (i.e. the set of all states) that are distinguishable from *x*. If *X* is a variable, Principle **IV** allows us to assign the natural meaning $\bar{X} \equiv \overline{\bigcup_{x \in X} x}$. (Thus any expression topped by a bar denotes an attribute.) When $\bar{X}$ is empty, we call *X* a *maximal variable*.

The bar operation has the property $\bar{\bar{\bar{x}}} \equiv \bar{x}$. For we have $x \subseteq \bar{\bar{x}}$ (since distinguishability is symmetric), i.e. $\bar{\bar{x}} = x \cup y$ for some *y*. Similarly, $\bar{\bar{\bar{x}}} = \bar{x} \cup z$ for some *z* that does not overlap with $\bar{x}$. Since, by definition, *z* contains only states distinguishable from $\bar{\bar{x}}$; and since $x \subseteq \bar{\bar{x}}$, each such state is also distinguishable from *x*. Hence *z* is empty and $\bar{\bar{\bar{x}}} \equiv \bar{x}$.

Any variable of the form $\{x,\bar{x}\}$ we shall call a *Boolean variable*. Principle **IV** and the definition of bar trivially imply that every Boolean variable is distinguishable[1]. Also, every Boolean variable is maximal, since no attribute $y$ can be distinguishable from both $x$ and $\bar{x}$, for if $y \perp x$, we would have $y \subseteq \bar{x}$ and hence $y \not\perp \bar{x}$.

Now consider an attribute $\{a\}$ in which $X$ is non-sharp. A trivial case is when $\{a\} \subseteq \bar{X}$; in that case, by principle **IV**, there is a distinguishable variable that includes $X$ that is sharp. But suppose instead that $\{a\} \subseteq \bar{\bar{X}}$, so that $\{a\} \not\subset \bar{X}$ (which includes all cases when $X$ maximal, because then $\bar{\bar{X}}$ includes all states). Whether it is then still possible for $X$ not to be sharp in the state $a$, and what it means if that is so, is up to the subsidiary theories. (For instance, in quantum theory it would mean that $a$ was a superposition or mixture of the states with attributes in $X$. The generalisation of this to constructor theory will be discussed in Section 8.5.) But it must at least imply that the output variable of every measurer of $X$ is either non-sharp, or sharp with some value '$x$' where $x \in X$. That means that the measurer could mistake the attribute $\{a\}$ for one having an attribute in $X$. For if some measurer of $X$ *could not* make such a mistake, it could distinguish $\{a\}$ from all attributes in $X$, contradicting $\{a\} \not\subset \bar{X}$.

Consider the case when a particular measurer of $X$ is presented with $\{a\} \subseteq \bar{\bar{X}}$ and the output variable is sharp with, say, the value '$x$'. That means that that measurer can mistake $\{a\}$ for only one attribute in $X$, namely $x$. Thus it is also a measurer of another variable, namely the variant of $X$ with the attribute $x$ replaced by $x \cup \{a\}$. Similarly we can construct a variable $Z$ by augmenting each attribute $x \in X$ with *all* the states $a \in \bar{\bar{X}}$ that can be mistaken only for $x$ by that measurer. Thus we see that the measurer was really a measurer of $Z$ all along, with the additional property that whenever it produces any sharp output '$z$' on measuring its substrate, the input substrate really had the attribute $z$. It is logically possible that applying this procedure for some other measurer of *X* would yield a different *Z*. But we propose the following *principle of the consistency of measurement*:

> **VIII.** Whenever a measurer of a variable *X* would produce a sharp output when presented with the attribute $\{a\} \subseteq \overline{\overline{X}}$, all other measurers of *X* would too[1].

---

[1] For the sake of uniformity of notation we include cases where $\bar{x}$ is empty, even though $\{x,\bar{x}\}$ is not literally *variable* in those cases. But since the empty set is an attribute that no substrate can possibly have, every state is distinguishable from it; hence in such cases $\bar{x}$ is the set of all states. A trivial constructor that produces a fixed result, which can then be labelled '$x$' in (5) then qualifies as a distinguisher of $\{x,\bar{x}\}$.

Consequently those other measurers of *X* would have to produce the *same* sharp output '*x*' for the given input $\{a\} \subseteq \overline{\overline{X}}$. For suppose one of them produced an output '*x*' for that input. That would make $\{a\}$ distinguishable from all attributes in *X* other than $x$. Likewise if another produced a different sharp output '*x*''. So then $\{a\}$ would be distinguishable from all $x \in X$, and would therefore (from **IV**) be included in $\overline{X}$ and hence not in $\overline{\overline{X}}$, which is a contradiction.

Thus, principle **VIII** implies that for any measurable variable *X* there is a unique variable *Z* such that all measurers of *X* are measurers of *Z*. *Z* therefore has the property that whenever a measurer of *Z* produces a sharp output '*z*' the input substrate really has the attribute *z*. We shall call such variables *observables,* and *information observables* if they are information variables. (Quantum-mechanical observables, in the Heisenberg picture, are indeed observables by this definition.) Properties of observables are crucial to our results about superinformation (Section 8).

*Properties of observables*

We now obtain a necessary and sufficient condition for a variable to be an observable. Consider *any* attribute *x*, and *any* measurable variable *X* of which it is a member. Now let $\chi$ be the union of all attributes in *X*, and consider the variable $X' = \{x, \chi - x\}$, a coarsening of *X*. It must be measurable since any measurer of $\{x, \overline{x}\}$ measures it. In this section and in Section 8.5 we shall repeatedly rely on the relations between measurers of *X*, $X'$, and the two Boolean variables $\{x, \overline{x}\}$ and $\{\overline{\overline{x}}, \overline{x}\}$, which are represented in Fig. 1.

---

[1] Despite mentioning constructors (measurers) explicitly, **VIII** is a purely constructor-theoretic statement, f<u>or</u> it could be rephrased as 'if the task of measuring *X* while producing an output '*x*' ($x \in X$) for an input $\{a\} \subseteq \overline{\overline{X}}$ is possible, the task of measuring *X* while not producing that output for the input $\{a\}$ is impossible'.

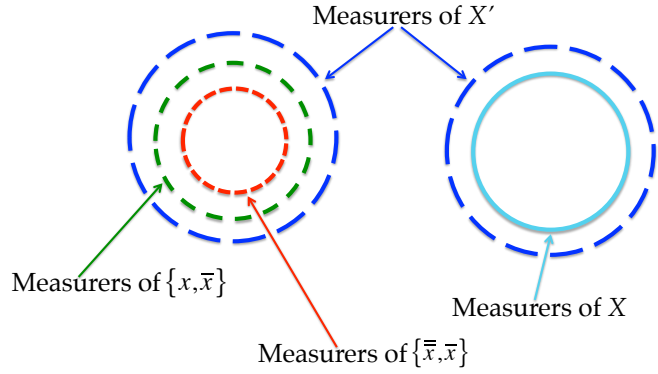

**Fig. 1 Hierarchy of measurers of variables of which *x* is a member.**

First we prove that all such measurers must produce a sharp '*x*' when presented with a substrate with the attribute $\{a\} \subseteq \bar{\bar{x}}$. By **VIII**, every measurer of $\{x, \bar{x}\}$ must deliver a sharp output '*x*' for any input $\{a\} \subseteq \bar{\bar{x}}$, since all measurers of $\{\bar{\bar{x}}, \bar{x}\}$ are measurers of $\{x, \bar{x}\}$ and would deliver a sharp '$\bar{\bar{x}}$', which could be reinterpreted as '*x*'. Any measurer of $\{x, \bar{x}\}$ measures X', and since the former produces a sharp output *x* when presented with $\{a\} \subseteq \bar{\bar{x}}$, by principle **VIII**, so do all measurers of X'. Also, any measurer of X is a measurer of X'. Therefore, in particular, any measurer of X must give a sharp output '*x*', when presented with any input $\{a\} \subseteq \bar{\bar{x}}$.

The converse is also true: if any measurer of a variable X (containing *x*) produces a sharp '*x*' when presented with a state $\{a\} \subseteq \bar{\bar{X}}$, then $\{a\} \subseteq \bar{\bar{x}}$. For suppose that given $\{a\} \in \bar{\bar{X}}$, a measurer of X produces a sharp outcome '*x*'. Then, all of them produce a sharp '*x*', by **VIII**. Again, any measurer of X is a measurer of X'; therefore, by **VIII**, all measurers of X' must give the same sharp output '*x*' when presented with $\{a\} \subseteq \bar{\bar{X}}$. By definition of $\bar{x}$, any measurer of $\{x, \bar{x}\}$ is a measurer of X'. Therefore, again by **VIII,** all measurers of $\{x, \bar{x}\}$ must give the sharp output '*x*' when presented with $\{a\} \in \bar{\bar{X}}$. This implies that $\{a\} \perp x$, i.e., $\{a\} \subseteq \bar{\bar{x}}$.

This elucidates the physical meaning of $\bar{\bar{x}}$: it is the set of all states *a* such that, for any variable X containing the attribute *x*, a measurer of X produces a sharp output '*x*' when presented with a substrate with the attribute $\{a\} \subseteq \bar{\bar{X}}$. And it also provides our necessary and sufficient condition for a variable to be an observable, namely: all its attributes $x$ satisfy $x = \bar{\bar{x}}$.

*Prediction, testing and ensembles*

Under constructor theory, a *testable prediction* is a statement that the output variable of a certain possible measurement will be sharp, with a certain value. That requirement of sharpness does not come from constructor theory. It is entailed by the logic of testability, and is why probabilistic predictions have to be tested by repeated measurements. For example, consider the testing of a fundamentally stochastic (and therefore incompatible with constructor theory) law of motion, which predicts the *probability* of an outcome of a measurement but not the outcome. Since probabilities of individual outcomes are not measurable on individual systems, such a prediction can be tested only if it is interpreted as a manner of speaking about non-probabilistic predictions about measurements on ensembles, which can then be approximated by average results of finite sets of measurements on individual systems.

Likewise, the constructor theory of measurement and testing must concern itself with how to obtain sharp outputs in situations where the individual outputs of measurements are not sharp, either because a subsidiary theory predicts that, or at a practical level because perfectly accurate measurers do not exist. And again, most methods of doing that involve using multiple instances of a given substrate.

We denote by $\mathbf{S}^{(n)}$ a physical system $\overbrace{\mathbf{S} \oplus \mathbf{S} \oplus \ldots \mathbf{S}}^{n \text{ instances}}$ consisting of $n$ instances of $\mathbf{S}$, and we denote by $x^{(n)}$ the attribute $\overbrace{(x, x, \ldots x)}^{n \text{ terms}}$ of $\mathbf{S}^{(n)}$. Evidently the set $\{x^{(n)} | x \in S\}$ is an information variable of $\mathbf{S}^{(n)}$ whenever $S$ is an information variable of $\mathbf{S}$. And each $x^{(n)}$ is a redundant instantiation of the information $x$.

If, in conventional, probabilistic, information theory, $S$ is a discrete set and each instance of $\mathbf{S}$ has a probability $p < \frac{1}{2}$ of having been changed from the correct value $x$ by uncorrelated perturbations, the probability that a plurality of the values stored in $n$ instances will not instantiate $x$ falls exponentially with $n$. Although probabilities are not allowed at the fundamental level in constructor theory, redundancy can still play effectively the same role, in following way. Denote by $\mathbf{S}^{(\infty)}$ an unlimited supply of instances of $\mathbf{S}$. Suppose that they all have the same intrinsic, preparable attribute, either $x$ or $y$, and let us call a sequence of experiments on $\mathbf{S}^{(n)}$, as $n$ increases without limit, an 'experiment on an ensemble $\mathbf{S}^{(\infty)}$'. Recall from Section 2 that a possible task is one for which the laws of nature impose no limit, short of perfection, on how well the task can be performed. Since our assumption **VI** implies that there is no limit on $n$, it follows that for any two intrinsic attributes $x$ and $y$ of a substrate $\mathbf{S}$, $x^{(\infty)}$ and $y^{(\infty)}$ are either interchangeable in all experiments on $\mathbf{S}^{(\infty)}$ (using generic resources), or distinguishable. For given such a supply, performing *all* possible experiments on (instances of) $\mathbf{S}$, infinitely often, is a possible task. So if that cannot tell the difference between $x^{(\infty)}$ and $y^{(\infty)}$, nothing (that uses only generic resources) can. If something can, we call $x$ and $y$ *ensemble distinguishable*. Thus we propose the principle:

   **IX.** Any two disjoint, intrinsic attributes are ensemble distinguishable.

## 7 Superinformation

So far in this paper we have proposed a set of purely constructor-theoretic principles that capture in an exact, subsidiary-theory-independent way the behaviour of 'classical' information. From here on, we propose no further principles. We investigate what happens if the subsidiary theories impose a single further *prohibition* on what tasks are possible. This turns out to allow substrates to instantiate what we call *superinformation*, and we shall show that quantum information is an instance of it.

It follows from our theory so far that every subset with at least two members of an information observable $S$, is also an information observable. But the converse does not hold: the union of two information observables, even if their attributes are mutually disjoint, is not necessarily an information observable. It is through that loophole that all the non-classical content of constructor information theory flows.

A *superinformation medium* **M** is an information medium with at least two information observables that contain only mutually disjoint attributes and whose union is not an information observable. For example, in quantum physics any set of two orthogonal states of a qubit constitutes an information observable, but no union of two or more such sets does: its members are not all distinguishable. **M** instantiates *superinformation* if it has one of the attributes in $S$ but could have had any of the others. A *supercomputation* is a task that maps $S$ to itself.

## 8 Properties of superinformation

In what follows, let **S** be a superinformation medium and let $X$ and $Y$ be two of its information observables whose union $X \cup Y$ consists of mutually disjoint attributes but is not an information observable. (Hence $X \cup Y$ is not even an information variable.)

*8.1 Not all information attributes of a superinformation medium are distinguishable*

There must exist information attributes $x \in X$ and $y \in Y$ such that $x \not\perp y$. For suppose that $x \perp y$ for all $x \in X$ and $y \in Y$. This would imply that all attributes in $X \cup Y$ are pairwise distinguishable and hence, from principle **IV**, that $X \cup Y$ is a measurable variable. Each attribute in $X \cup Y$ is also preparable, since by **VI** that is true of $X$ and $Y$ separately. It would follow that all permutations on $X \cup Y$ are possible tasks. For, to perform $\Pi$, given **S** with any attribute $z \in X \cup Y$, one would first measure which attribute that is, thereby preparing some information medium with the information attribute *'z'*. Then one would compute $\Pi('z')$ on that medium (which must be possible because it is a computation medium). Then one would use that result to prepare, from generic substrates, another instance of **S** with the attribute $\Pi(z)$, which again must be possible by principle **VI**.

The remaining condition for $X \cup Y$ to be an information variable would be met too: the cloning task is possible (with side-effects). For if $X \cup Y$ is a distinguishable set of **S**, $(X \cup Y) \times (X \cup Y)$ is a distinguishable set of $\mathbf{S} \oplus \mathbf{S}$: one can distinguish its members by performing a distinguishing operation on each instance of S in parallel and then combining the sharp outputs with a logical-or operation. Moreover, $X \times X$, $X \times Y$, $Y \times X$ and $Y \times Y$, are all information variables of $\mathbf{S} \oplus \mathbf{S}$, by the interoperability principle **III**. Therefore each attribute in their union $(X \cup Y) \times (X \cup Y)$ is preparable. These two facts imply, by the argument above, that all permutations of $(X \cup Y) \times (X \cup Y)$ are possible with side-effects. Since those permutations include the cloning tasks on $X \cup Y$, it follows that $X \cup Y$ is an information variable, which contradicts the condition for **S** to be a superinformation medium.

So there must exist a pair of attributes $x \in X$ and $y \in Y$ of **S** that are not distinguishable.

*8.2 Undetectability of sharpness*

It is impossible to measure whether the observable *X* or *Y* is sharp. For it were possible, that would also distinguish between the above-mentioned *x* and *y*. (Of course sharpness is ensemble measurable, by **IX**, because none of the attributes in *X* intersects any attribute in *Y*.)

*8.3 Superinformation cannot be cloned*

Suppose that the cloning task (3) were possible for the variable $X \cup Y$. Then in particular, any variable $\{x,y\}$ with $x \in X$ and $y \in Y$ could be cloned. Then, given our generic resources assumption **VII**, if $z \in \{x,y\}$ it is possible to apply the cloning operation to the substrate any number of times, and the output would be a composite medium $\mathbf{S} \oplus \mathbf{S} \oplus \mathbf{S}\ldots$ with the attribute $(z,z,z\ldots)$. Thus, preparing $z^{(\infty)}$ would be a possible task. The attributes *x* and *y* are intrinsic information attributes, so by assumption **VI** they are preparable, and therefore by principle **IX** the two attributes $z^{(\infty)}$ are ensemble distinguishable: $x^{(\infty)} \perp y^{(\infty)}$. Thus by preparing $z^{(\infty)}$ from $\{x,y\}$ one could distinguish *x* from *y*.

If all such pairs were clonable, it would follow that all attributes in $X \cup Y$ were pairwise distinguishable, contrary to our result in 8.1, so the assumption that the superinformation variable $X \cup Y$ is clonable is false.

*8.4 Pairs of observables not simultaneously preparable or measurable*

The sets $x \cap y$ are all empty by the defining property of a superinformation medium. So when the substrate has any of the attributes $y \in Y$, $X$ cannot be sharp, and vice versa; hence it is impossible to prepare **S** with its observables *X* and *Y* both sharp. For the same reason, simultaneously measuring *X* and *Y* is impossible.

*8.5 Unpredictability of deterministic processes*

Superinformation media exhibit the counter-intuitive property of evolving deterministically yet unpredictably.

Unpredictability arises when a measurer of *X* acts when *X* is non-sharp. Suppose that *X* is maximal, whereby $y \subseteq \overline{\overline{X}}$. If a substrate with attribute *y* is presented to a measurer of *X*, the output variable cannot be sharp. For if it were, with value '*x*', the properties of observables (Section 6), would imply that $y \subseteq x$; yet *x* and *y* are non-overlapping by hypothesis. Thus, no prediction of the form «the outcome will be '*x*'», where $x \in X$, can be true, since that would imply that the measurer, when presented with *y*, produces a sharp output. That already means that the outcome of a measurement of *X*, when the substrate has the attribute *y*, is unpredictable. Remarkably, constructor information theory also provides a definite physical meaning for this unpredictability, even though the relevant subsidiary theories must give no meaning to probabilities.

As a guide to the general constructor-theoretic case, let us consider an example from quantum theory. An observable such as *X* might be the number of photons in a cavity, $\hat{N} = |1\rangle\langle1| + 2|2\rangle\langle2| + 3|3\rangle\langle3| + \ldots$ The only way that $\hat{N}$ can fail to be sharp when the photon field in the cavity has some attribute *y* is that *y* contains superpositions or mixtures of two or more eigenstates of $\hat{N}$ – let us say, those with eigenvalue less than *3*. In such a case, when $\hat{N}$ is measured, no prediction of the form «the outcome will be '$n$'» (for some eigenvalue $n$ of $\hat{N}$) would be true. But there are other predictions that must be true: any pure state $|\psi\rangle$ in $y$ has the property that the expectation value of the projector $|0\rangle\langle 0|+|1\rangle\langle 1|+|2\rangle\langle 2|$ is 1 in $|\psi\rangle$. Therefore, the prediction that a measurer of $|0\rangle\langle 0|+|1\rangle\langle 1|+|2\rangle\langle 2|$ would yield the outcome '1', when presented with $|\psi\rangle$ would be true. Thus, since $|0\rangle\langle 0|+|1\rangle\langle 1|+|2\rangle\langle 2|$ is a Boolean observable whose meaning is «*whether there are fewer than 3 photons in the cavity*» (with 1 denoting yes and 0 no), the fact that the outcome of a measurement of $\hat{N}$ would be less than 3 *is* predictable, even though predicting the outcome itself is an impossible task. This is the physical meaning of unpredictability in the quantum case.

The same logic applies equally in constructor-theory, as we shall now explain, without the apparatus of projectors, expectation values and probabilities.

Let $\chi$ again be the union of all attributes in $X$ and $\chi_y$ be the union of all attributes in $X$ that are not distinguishable from $y$. (By the defining properties of superinformation, there must exist at least one such attribute.) Our goal is now to explain that the Boolean observable $\{\overline{\overline{\chi_y}},\overline{\chi_y}\}$ plays the same role as the above projector. We start by showing that it must be sharp with value $\overline{\overline{\chi_y}}$ when the substrate has the attribute $y$.

Consider any state $a \in y$. Any measurer of the Boolean variable $\{\chi-\chi_y, \overline{\chi-\chi_y}\}$, when presented with a substrate with the attribute $\{a\}$, will produce a sharp output '$\overline{\chi-\chi_y}$' because $y \perp \chi-\chi_y$ (by definition of $\chi_y$). Any measurer of $\{\chi-\chi_y, \overline{\chi-\chi_y}\}$ is also a measurer of the (maximal) variable $\{\chi-\chi_y, \chi_y\}$, (because $\chi_y \subseteq \overline{\chi-\chi_y}$). Hence by principle **VIII** all measurers of $\{\chi_y, \chi-\chi_y\}$ must give a sharp output '$\chi_y$' when presented with $\{a\} \subseteq y$. By the properties of observables (Section 6), we conclude that $\{a\} \subseteq \overline{\overline{\chi_y}}$. Since this is true for any state in $y$, it must be true of $y$ too: $y \subseteq \overline{\overline{\chi_y}}$. Therefore, again by the result of Section 6, when the attribute is $y$, $\{\overline{\overline{\chi_y}},\overline{\chi_y}\}$ must be sharp with value $\overline{\overline{\chi_y}}$.

Consider now the observable $X_y = \{x \in X : x \not\sqsubseteq y\}$ (noting that $\overline{\overline{\chi_y}} = \overline{\overline{X}}_y$). First, it must contain at least two attributes. For suppose all the attributes in $X$ except $x$ were distinguishable from $y$. Then $\chi_y = x$, so that $y$ would be included in $\overline{\overline{x}}$. But this is a contradiction, because this would imply that $y \subseteq x$ (since $X$ is an observable).

We can now see why $\{\overline{\overline{\chi_y}}, \overline{\chi_y}\}$ generalises the projector in the quantum example. One way of measuring $\{\chi_y, \overline{\chi_y}\}$ is to measure $X$ first and then to perform a computation on the output '$x$' that would, if $X$ were sharp in the input, determine whether $x \in X_y$ or not. All such measurers must, by the principle of consistency of measurement **VIII**, give a sharp '$\chi_y$' when presented with any attribute in $\overline{\overline{\chi_y}}$; hence they are all measurers of $\{\overline{\overline{\chi_y}}, \overline{\chi_y}\}$ too. Thus $\{\overline{\overline{\chi_y}}, \overline{\chi_y}\}$ is a Boolean observable whose meaning is «whether the outcome is one of the '$x$' with $x \not\sqsubseteq y$». Since $y \subseteq \overline{\overline{\chi_y}}$, that process, by principle **VIII**, must yield the same sharp output '$\overline{\overline{\chi_y}}$', meaning 'yes', as any other measurer of $\{\overline{\overline{\chi_y}}, \overline{\chi_y}\}$ does when presented with $y$. As in the quantum case, this provides the physical meaning of unpredictability: Any constructor (including any observer) that measures $X$ on a substrate with attribute $y$ and then computes (or recollects) whether the outcome was one of the '$x$' with $x \not\sqsubseteq y$, by the above procedure, will reach the conclusion 'yes' (corresponding to '$\overline{\overline{\chi_y}}$'); And will thereby have the same attribute as it would have if $X$ had been sharp with some value $x \in X_y$. Yet no prediction «the outcome will be '$x$'» with $x \in X$ will be true.

Again, it is up to the subsidiary theories to *explain* this deterministic unpredictability. In Everettian quantum theory, the explanation is that the measurer differentiates, during the measurement, into multiple instances, sharply agreeing that the output was one of the $x \in X, x \not\sqsubseteq y$ but not all perceiving the same one.

*8.6 Irreducible perturbation of one observable caused by measuring another*

The observable $X_y$ contains two or more attributes, none of which is distinguishable from $y$ nor overlaps with $y$. We shall now show that any measurer of $X_y$ must cause an irreducible perturbation of the substrate for some input attributes. In particular:

$$\left( \bigcup_{x \in X_y} \{(x, x_0) \rightarrow (x, 'x')\} \ \bigcup \ \{(y, x_0) \rightarrow (y, k)\} \right)^{\boldsymbol{x}} \tag{6}$$

for all $x_0$ and $k$. In words: no device can both measure $X_y$ non-perturbatively if $X_y$ is sharp and leave the substrate unperturbed if the input has attribute $y$. Moreover, the perturbation is irreducible: nothing can subsequently undo it while leaving the outcome of the measurement in any information variable, for if it could, the overall process would be a counter-example to (6).

To prove (6), suppose that the task in (6) were possible. Since $y$ is an information attribute (hence, it is intrinsic) and $x_0$ is preparable from generic resources, $k$ would be an intrinsic attribute because, by hypothesis, it could be produced from generic resources in the combination $(y, k)$ with the intrinsic attribute $y$.

Moreover, $k$ would be disjoint from all the '$x$'. For suppose there were a state in $k \cap {}'x'$, for some '$x$'. That would imply that there exists a state $a$ in $y$ with the property that $\{a\}$, when presented to any measurer of $X_y$, produced an output with value '$x$'. Thus, by the results in Section 6, it would follow that $\{a\} \subseteq x$ (because $X_y$ is an observable), but this is a contradiction, as $y$ does not intersect any $x$.

Therefore $k$ would be ensemble distinguishable from $x$. Thereby it would be possible, by performing the above task an unlimited number of times on the same instance of the substrate together with successive instances of the target with attribute $x_0$, to produce an ensemble with attribute $k^{(\infty)}$ or ${}'x'^{(\infty)}$, and so $k$ would be distinguishable from '$x$', and so $y$ would be distinguishable from $x$, contradicting the supposition.

*8.7 Consistency of consecutive measurements of a non-sharp observable*

In quantum theory, consecutive non-perturbing measurements of an observable, even if it is not sharp, yield the same outcomes, in the sense that a measurement of whether they are the same always yields the sharp outcome 'true'. We now show that this also holds for arbitrary subsidiary theories that conform to constructor theory.

A somewhat roundabout way of measuring the variable $\{\chi_y, \overline{\chi_y}\}$ of **S** is the following. First, apply two non-perturbing measurers of $X$ in succession to the substrate, recording each output in one of two receptive information media **M** and **M'**. Then present **M** to a computer that performs

$$\bigcup_{x \in X_y} \{x \to x\} \ \cup \ \bigcup_{x \in (X - X_y)} \{x \to \Pi(x)\} \tag{7}$$

where $\Pi$ is any permutation with no fixed point. (Here we have suppressed the quotation marks to avoid clutter, replacing '$x$' by $x$, writing $X$ for the output variable of **M**, and $X'$ for that of **M'**, and so on.) This leaves the attributes in $X_y$ (of **M**) unchanged and changes those in $X - X_y$ (of **M**). Finally, present **M**, **M'** and a third, receptive, information medium **R** to a computer that performs

$$\bigcup_{v \in X, w' \in X'} \{(v, w', x_0) \to (v, w', v = w'?)\}. \tag{8}$$

The question mark means that when $X$ of **M** and $X'$ of **M'** are sharp, the output observable $\{\text{'true', 'false'}\}$ of **R** indicates whether their values $v$ and $w'$ are equal or not.

Call the combined measurer that performs this overall process **N**. It is also a measurer of $\{\chi - \chi_y, \chi_y\}$ because when **S** has attribute $x \in X_y$, the output observable of **R** is sharp with the value 'true', which can also be interpreted as '$\chi_y$'; and similarly when **S** has attribute $x \in X - X_y$, it is sharp with the value 'false', meaning '$\chi - \chi_y$'. By principle **VIII**, **N** is also a measurer of $\{\chi_y, \overline{\chi_y}\}$.

Now suppose that **N** is presented with a substrate with the attribute $y$. Since $y \subseteq \overline{\overline{\chi_y}}$ (Section 8.5), by principle **VIII** the measurer **N** must deliver a sharp outcome '$\chi_y$' (i.e. 'true'), which is the same as it would deliver if the outputs of the two consecutive measurements of $X$ were both sharp with the same value – even though, as we have seen, neither of them can be sharp when $y$ is the input, just as in quantum theory.

*8.8 Quantisation*

The states of a quantum system form a continuum, and their dynamical evolution is continuous in both space and time. With hindsight we can now see that the 'quantisation' after which quantum theory is named really refers to a property of quantum *information.* The discrete and the continuous are linked, in quantum theory, in a manner that was not previously guessed at, but is easily understood in terms of constructor information theory: each information observable of a quantum physical system has only a discrete set of attributes, but there is a continuous infinity of such observables, no union of which is an information observable. So in quantum physics, classical information is discrete, and superinformation (quantum information) is continuous.

*8.9 Coherence and locally inaccessible information*

Another feature of quantum theory that is due to its permitting superinformation is the distinction between coherent and incoherent processes. Let $w$ be a set of states of a superinformation medium **M**. A computation or supercomputation $\mathbb{C}$ on a proper subset $v$ of $w$ is *coherent* with respect to $w$ if it is possible to perform it reversibly on $w$. That is to say, there exists a task $\mathfrak{A}$, with $\mathfrak{A}^{\checkmark}$ and $\mathfrak{A}^{\sim\checkmark}$, whose legitimate input set contains $w$ and whose restriction to the subtask with legitimate input set $v$ is $\mathbb{C}$. **M** is a *coherent medium* with respect to information variables $\{S_i\}$ of **M** if each permutation task on each $S_i$ can be performed coherently with respect to the union of all the attributes in the $\{S_i\}$.

For example, the qubits of a universal quantum computer constitute a coherent superinformation medium, because they are a superinformation medium and all reversible classical computations in some computation basis can be performed reversibly on the set of all their pure states.

Quantum entanglement is an example of a phenomenon that depends on coherence. It is usually characterised in terms of probabilistic quantities such as the correlations referred to in Bell's theorem. But underlying those quantitative measures is a qualitative property:

the presence of *locally inaccessible information*[1] (Deutsch & Hayden 2000). In our terminology, that means that some combined system $\mathbf{S}_1 \oplus \mathbf{S}_2$ has information variables that cannot be measured by measuring Cartesian products (nor subsets thereof) of variables of $\mathbf{S}_1$ and $\mathbf{S}_2$.

So, let $\mathbf{S}_1$ be a superinformation medium, as in Section 7. Again let $X$ be maximal. Consider any twofold observable $B_1 = \{0', 1'\} \subseteq Y$. We proved in Section 8.5 that for any attribute $y \in Y$ there are at least two attributes in $X$ which are not distinguishable from $y$. Let $A_1 = \{0, 1\} \subseteq X$ be the observable including two attributes that are not distinguishable from $0'$, and define $A_2$ and $B_2$ for $\mathbf{S}_2$ analogously. Now suppose that $\mathbf{S}_1 \oplus \mathbf{S}_2$ is coherent with respect to the variables $\{A_1 \times A_2, B_1 \times A_2\}$. These are information observables, so that in particular we have $\mathcal{T}^{\swarrow}$ where

$$\mathcal{T} = \{(0,0) \to (0,0),\ (1,0) \to (1,1),\ (0,1) \to (0,1),\ (1,1) \to (1,0)\}. \tag{9}$$

$\mathcal{T}$ is the controlled-not computation. The coherence condition implies that in addition, $(\mathcal{T} \cup \mathcal{T}')^{\swarrow}$ and $(\mathcal{T} \cup \mathcal{T}')^{\widetilde{\swarrow}}$, where

$$\mathcal{T}' = \{(0',0) \to \psi_{(1)},\ (0',1) \to \psi_{(2)},\ (1',0) \to \psi_{(3)},\ (1',1) \to \psi_{(4)}\}, \tag{10}$$

and $\psi_{(1)} \ldots \psi_{(4)}$ are four distinct states of $\mathbf{S}_1 \oplus \mathbf{S}_2$. Because of the principle of locality **II**, these must have the ordered-pair form $\psi_{(1)} = (a_{(1)}, b_{(1)})$, etc., where $a_{(1)} \ldots a_{(4)}$ are attributes (not necessarily intrinsic) of $\mathbf{S}_1$ and $b_{(1)} \ldots b_{(4)}$ of $\mathbf{S}_2$.

Now, $C = \{(0,1), (1,1), (0',0), (1',0)\}$ is an information variable. (Evidently one can measure it by first measuring $A_2$ and then either $A_1$ or $B_1$ according to the output of that measurement, which is necessarily sharp.)

The effect of performing the task $\mathcal{T} \cup \mathcal{T}'$ on the substrate when $C$ is sharp must be to make the variable $D = \{(0,1), (1,0), \psi_{(1)}, \psi_{(3)}\}$ sharp with the corresponding values under

---

[1] Entanglement is not the only property of quantum information for which locally inaccessible information is responsible. There is also the misleadingly named "non-locality without entanglement" of Bennett *et al.* (1999).

(9) and (10). And *D* must also be an information variable because, by the coherence property, $(\mathcal{C} \cup \mathcal{C}')^\sim$ is also possible.

Now consider the subset $\{(0,1), \psi_{(1)}\}$ of *D*. This is an information variable that cannot be measured by measuring products of variables of $\mathbf{S}_1$ and $\mathbf{S}_2$. For if it could, then $a_1 \perp 0$ or $b_1 \perp 1$. But if $a_1 \perp 0$ then $\psi_1 \perp (0,0)$, which in turn would imply that $(0',0) \perp (0,0)$ and hence that $0' \perp 0$, contrary to construction. Similarly $b_1 \perp 1 \Rightarrow \psi_1 \perp (1,1) \Rightarrow (0',0) \perp (1,0) \Rightarrow 0' \perp 1$, again contrary to construction.

So the information variable $\{(0,1), \psi_{(1)}\}$ holds locally inaccessible information, as promised. Note that the failure of local distinguishability is possible because superinformation media *obey* the principle of locality **II**, not disobey it as has been supposed.

**9 Concluding remarks**

The constructor theory of information relies only on the fundamental constructor-theoretic dichotomy between possible and impossible tasks. All its definitions and conjectured principles are constructor-theoretic. It reconciles apparently contradictory features of information: that of being an abstraction, yet governed by laws of physics; of being physical, yet counter-factual. And it robustly unifies the theories of quantum and classical information.

**Acknowledgments**

We thank Harvey Brown, Oscar Dahlsten and Alan Forrester for suggesting improvements to earlier drafts of this paper. This work was supported by the Templeton World Charity Foundation.